# Data-Driven Key Performance Indicators and Datasets for Building Energy Flexibility: A Review and Perspectives

Han Li [1], Hicham Johra [2,*], Flavia de Andrade Pereira [3,4], Tianzhen Hong [1], Jérôme Le Dréau [5], Anthony Maturo [6], Mingjun Wei [7], Yapan Liu [8], Ali Saberi-Derakhtenjani [9], Zoltan Nagy [10], Anna Marszal-Pomianowska [2], Donal Finn [3], Shohei Miyata [11], Kathryn Kaspar [12], Kingsley Nweye [10], Zheng O'Neill [7], Fabiano Pallonetto [13], Bing Dong [8]

[1] *Lawrence Berkeley National Laboratory, Building Technology and Urban Systems Division, Berkeley, CA, USA.*
[2] *Aalborg University, Department of the Built Environment, Thomas Manns Vej 23, DK-9220 Aalborg Ø, Denmark.*
[3] *University College Dublin, Energy Institute, School of Mechanical & Materials Engineering, Dublin, Ireland.*
[4] *CARTIF Technology Center, Energy Division, Valladolid, Spain.*
[5] *La Rochelle University, LaSIE UMR CNRS 7356, La Rochelle, France.*
[6] *Concordia University, Centre for Zero Energy Building Studies, Montreal, Canada.*
[7] *Texas A&M University, J. Mike Walker '66 Department of Mechanical Engineering, College Station, TX, USA.*
[8] *Syracuse University, Department of Mechanical and Aerospace Engineering, 223 Link Hall, Syracuse, NY, 13244, USA.*
[9] *Dubai Electricity and Water Authority, DEWA R&D Centre, Dubai, United Arab Emirates.*
[10] *Department of Civil, Architectural and Environmental Engineering, University of Texas at Austin, Texas, USA.*
[11] *The University of Tokyo, Department of Architecture, Tokyo, Japan.*
[12] *Concordia University, Department of Building, Civil and Environmental Engineering, Montreal, QC, Canada.*
[13] *Maynooth University, Hamilton Institute, Maynooth, Ireland.*

*\* Corresponding author. Tel.: +45 5382 8835. E-mail address: hj@build.aau.dk (H. Johra).*

**Abstract**

Energy flexibility, through short-term demand-side management (DSM) and energy storage technologies, is now seen as a major key to balancing the fluctuating supply in different energy grids with the energy demand of buildings. This is especially important when considering the intermittent nature of ever-growing renewable energy production, as well as the increasing dynamics of electricity demand in buildings. This paper provides a holistic review of (1) data-driven energy flexibility key performance indicators (KPIs) for buildings in the operational phase and (2) open datasets that can be used for testing energy flexibility KPIs. The review identifies a total of 48 data-driven energy flexibility KPIs from 87 recent and relevant publications. These KPIs were categorized and analyzed according to their type, complexity, scope, key stakeholders, data requirement, baseline requirement, resolution, and popularity. Moreover, 330 building datasets were collected and evaluated. Of those, 16 were deemed adequate to feature building performing demand response or building-to-grid (B2G) services. The DSM strategy, building scope, grid type, control strategy, needed data features, and usability of these selected 16 datasets were analyzed. This review reveals future opportunities to address limitations in the existing literature: (1) developing new data-driven methodologies to specifically evaluate different energy flexibility strategies and B2G services of existing buildings; (2) developing baseline-free KPIs that could be calculated from easily accessible building sensors and meter data; (3) devoting non-engineering efforts to promote building energy flexibility, standardizing data-driven energy flexibility quantification and verification processes; and (4) curating and analyzing datasets with proper description for energy flexibility assessments.

**Highlights**

- Studies on data-driven methods for building energy flexibility quantification have been reviewed.
- Existing data-driven energy flexibility KPIs have been systematically categorized.
- Various aspects of the KPIs, including application, stakeholder, complexity, data requirements, and popularity have been analyzed.
- Public datasets for energy flexibility studies have been reviewed and summarized in a standardized data collection process.
- Research trends, open questions, and future opportunities are identified.

**Nomenclature**

| Acronyms | Definitions |
|---|---|
| ADR | Automated Demand Response |
| B2G | Building-to-Grid |
| DR | Demand Response |
| DSM | Demand-Side Management |
| DSO | Distribution System Operator |
| EF | Energy Flexibility (usually used interchangeably with Demand Flexibility) |
| EV | Electric Vehicle |
| GEB | Grid-interactive Efficient Building |
| GHG | Greenhouse Gas |
| HIL | Hardware-In-the-Loop |
| HVAC | Heating, Ventilation, and Air-Conditioning |
| IEA | International Energy Agency |
| IEA EBC | International Energy Agency - Energy in Buildings and Communities Programme |
| IEQ | Indoor Environmental Quality |
| IoT | Internet of Things |
| KPI | Key Performance Indicator |
| PV | Photovoltaics |
| TSO | Transmission System Operator |

## 1. Introduction

*1.1. General background*

The building sector's entire life cycle is directly or indirectly responsible for about 36% of the global primary energy demand and about 37% of energy-related carbon dioxide ($CO_2$) emissions (operational, embedded, and construction) [1][2]. Decarbonizing the building sector is thus essential to achieving a global carbon-neutral society by 2060 [3]. Meanwhile, another challenge to address is climate resilience, as more frequent, intense, and longer-lasting extreme weather events exacerbated by climate change, such as heat waves, occur. The energy resilience of buildings and energy grids have, therefore, become essential in providing critical cooling and heating services to occupants to avoid excessively high or low indoor temperatures or energy outages. The latter threaten the lives of citizens and cause illness and detrimental health consequences. Decarbonizing buildings and improving their climate resilience should be tackled together with evaluating technologies or design and operational strategies to improve building performance. Opportunities emerge with the penetration of digital technologies such as smart meters, indoor environmental quality (IEQ) sensors, and Internet of Things (IoT) devices, the use of which is expected to grow significantly in the building industry [4][5]. This growing smart building trend opens up new data sources feeding advanced analytic algorithms to inform the design and control of buildings.

With electrification of building energy demand (e.g., space heating/cooling, domestic hot water, cooking) becoming a key strategy to building decarbonization [6], there is growing dependence of building energy provision and resilience on the capacity and reliability of the energy grids. Energy flexibility, through demand-side management (DSM), demand response (DR), and energy storage technologies is increasingly seen as critical in balancing the electric power supply and demand for buildings, especially considering the intermittent nature of the growing renewable energy production from solar photovoltaics (PV) and wind generation, as well as increasing dynamics of electricity demand in buildings and electric vehicle (EV) charging. DSM also can be very beneficial for other energy grids, such as district heating/cooling (DHC) networks. DSM can help to decarbonize DHC by means of peak shaving, thus eliminating the use of $CO_2$-intensive auxiliary boilers, but also by lowering the temperature supply (towards energy-efficient fourth generation district heating) and tackling local bottleneck/power congestion problems. An international group of researchers under the umbrella of the IEA EBC[1] Annex 82 *Energy Flexible Buildings Towards Resilient Low Carbon Energy Systems* recently published an article identifying 10 key questions on the energy flexibility of buildings [7]. In their second question (How can energy flexibility be quantified?), they describe key performance indicators (KPIs) as essential to quantifying energy flexibility performance considering available flexible resources, building demand, grid signals, and occupant comfort needs or constraints.

*1.2. Building energy flexibility*

According to Al Dakheel et al. [8], smart buildings can be defined by four main features: (1) climate response, (2) grid response, (3) user response, and (4) monitoring and supervision. These types of buildings must react appropriately to external climate conditions, signals/information coming from the grid, and real-time interaction between users and implemented technologies, and must carry out real-time monitoring of a building's operation. With the upcoming challenges of increasing energy demand and renewable power generation, electrification penetration, and global warming, DR of the building stock becomes an increasingly important strategy for a safe and cost-effective energy system [9]. Buildings can provide grid services via flexible operations (e.g., adjusting their energy demand and behind-the-meter power generation and storage) [10]. An example of a flexibility service is the Automated Demand Response (ADR) event [11]. This allows electric devices to be turned off during periods of high demand via an internal control signal from the building control system, or direct external control by the grid operator, or indirect external control with an incentive grid signal, such as energy spot price or $CO_2$ intensity, to which the local controller reacts to trigger DR.

Energy flexibility in buildings has gained growing research momentum in recent years. A number of national and international collaborations and initiatives, including the IEA EBC Annex 67 [12], Annex 81 [13], Annex 82 [14], Annex 84 [15], as well as the GEB initiative by the U.S. Department of Energy (U.S. DOE) [16], have been trying to bring building energy flexibility to the next level of maturity. The IEA EBC Annex 67 developed a common definition of building energy flexibility as *"the ability to adapt/manage its short-term (a few hours or a couple of days) energy demand and generation according to local climate conditions, user needs, and energy network requirements without jeopardizing the technical capabilities of the operating systems in the building and the comfort of its occupants. Energy Flexibility of buildings will thus allow for DSM/load control and thereby DR based on the requirements of the surrounding energy grids"* [12]. Research on building energy flexibility-related topics is growing rapidly, including (1) equipment and system development, (2) modelling and simulation, (3) controls and energy system management, and (4) energy flexibility characterization and quantification.

*1.3. Motivations for data-driven energy flexibility KPIs*

While there are many existing studies on the energy flexibility of buildings, they usually rely on detailed building energy models that can simulate the baseline building operation and the flexible operation to calculate the energy flexibility by comparison of the two. However, most such studies lack real measured performance data to validate the modelled results. Moreover, developing and calibrating building energy models is time-consuming and requires expertise. Thus it is hard to scale up the deployment to many buildings. As Li et al. [7] suggested, building energy flexibility is not an invariant intrinsic parameter; it varies depending on the available resources and specific objectives and is constrained by occupants' needs. Therefore, energy flexibility quantification methods should allow real-time updates according to the performance goals. In this context, data-driven approaches are being used more and more to understand and quantify the energy flexibility of buildings with KPIs. These KPIs vary in definitions, amount and type of data required, and characteristics of the buildings and energy systems considered, especially with distributed energy sources such as on-site

---
[1] International Energy Agency - Energy in Buildings and Communities Programme

PV, energy storage, and EVs. A small portion of the existing KPIs can be applied directly to energy performance measurement data, while the majority requires comparison with a baseline or reference case, which can be very cumbersome to obtain. With the increasingly popular deployment of sensing and metering in buildings, an ever-growing amount of data is generated, which enables the adoption of data-driven approaches to compute energy flexibility KPIs. However, a holistic review of data-driven KPIs is lacking for the operational phase of commercial and residential buildings across scales (individual buildings, cluster of buildings, district). This can be of great interest for several stakeholders like building owners, households, building managers, utility companies, distribution system operators (DSOs), transmission system operators (TSOs), energy brokers, and market aggregators.

The above-mentioned gap motivated this article, which provides a review and insights into a few important aspects of KPIs for building energy flexibility assessment:

- **What are the use cases for energy flexibility KPIs?** KPIs play a key role in quantifying the energy flexibility of buildings [7]. However, due to the diversity of application scopes, technologies, data types, and interested stakeholders, there is no clear picture of when and how these KPIs can be used. This study systematically reviewed their associated application scopes, technologies, assessment methods, and targeted stakeholders.

- **What are the existing data-driven energy flexibility KPIs?** A small group of data-driven energy flexibility KPIs can be computed directly from building performance data without the need for baseline demand profiles. However, the majority of energy flexibility KPIs rely on the comparison between the building performance profiles under the baseline and flexible operation scenarios. The current study reviewed the existing publications that involve data-driven energy flexibility assessments and systematically categorized them into baseline-free and baseline-required groups.

- **What are available datasets for energy flexibility KPI development?** Building performance data is the foundation of data-driven energy flexibility assessments. Although there are many existing efforts in data curation for building performance modelling and evaluations [17][18][19][20], the datasets are not dedicated to building energy flexibility quantification. Moreover, there is no guidance on how existing datasets could be used for energy flexibility KPI development. In this paper, candidate datasets suitable for energy flexibility KPI development were surveyed and identified. The building characteristics, metadata, and detailed sensor and meter data format and quality for the candidate datasets were reviewed.

To provide comprehensive answers to the aforementioned questions, this review paper is organized as follows: Section 2 introduces the scope, objectives, and search methods of the literature review; Section 3 presents the review findings on use cases when data-driven energy flexibility KPIs can be applied or preferred; Section 4 focuses on existing data-driven energy flexibility KPIs; Section 5 provides a review of datasets for energy flexibility KPI development; Section 6 summarizes the key findings and contributions of the article, as well as identifying possible future work; and Section 7 closes with the main conclusions.

## 2. Methodology

*2.1. Scope of the KPI review*

As indicated above, the key limitations of data-driven energy flexibility quantification include: (1) the lack of data-driven methods to generate both baseline and flexible load profiles for KPI calculations (baseline-required KPIs) and (2) the lack of KPIs that do not require input from baseline or reference scenarios (baseline-free KPIs). Therefore, the literature search sought to cover the publications most relevant to the aforementioned challenges (see *Figure 1*).

Specifically, the review focuses on data-driven solutions for buildings during their operational phase. Data-driven solutions are here limited to methods that do not rely on pre-existing detailed models of the building (white box models), prior knowledge of the building's characteristics, or extensive human inputs. Data-driven approaches such as automated grey box/black box calibration/system identification for the generation of baselines or DR profiles are thus included. Human-input-intensive approaches requiring tailored white box models or specific on-site tests of the building case have low scalability and automation potential; they are thus deemed non-data-driven and are excluded from the scope of this review. Moreover, this study targets KPIs usable for energy flexibility assessment of the buildings in the operational phase. This adds certain constraints regarding data requirements and availability compared to the planning and design phases. Indeed, the latter usually rely on detailed models of the building from which far more data can be extracted. KPIs for planning and design phases only were thus excluded from this review. However, the energy flexibility KPIs for the operational phase can probably also be used for the design phase since all data of the operational phase can usually be generated or estimated in the design phase.

The KPIs within the scope of this review are thus primarily intended to be used with real monitoring data from existing buildings during the operational phase. However, many publications on energy flexibility used numerical models to generate DR energy profiles. If this numerical model was used as if it were an existing building (emulating the behaviour of an energy user performing DR but not directly generating comparison baseline profiles), the study was included in the review, as its methodology is data-driven and can be applied directly to a real-world case without the need to create a dedicated white box model. If not, the publication was excluded from the current review.

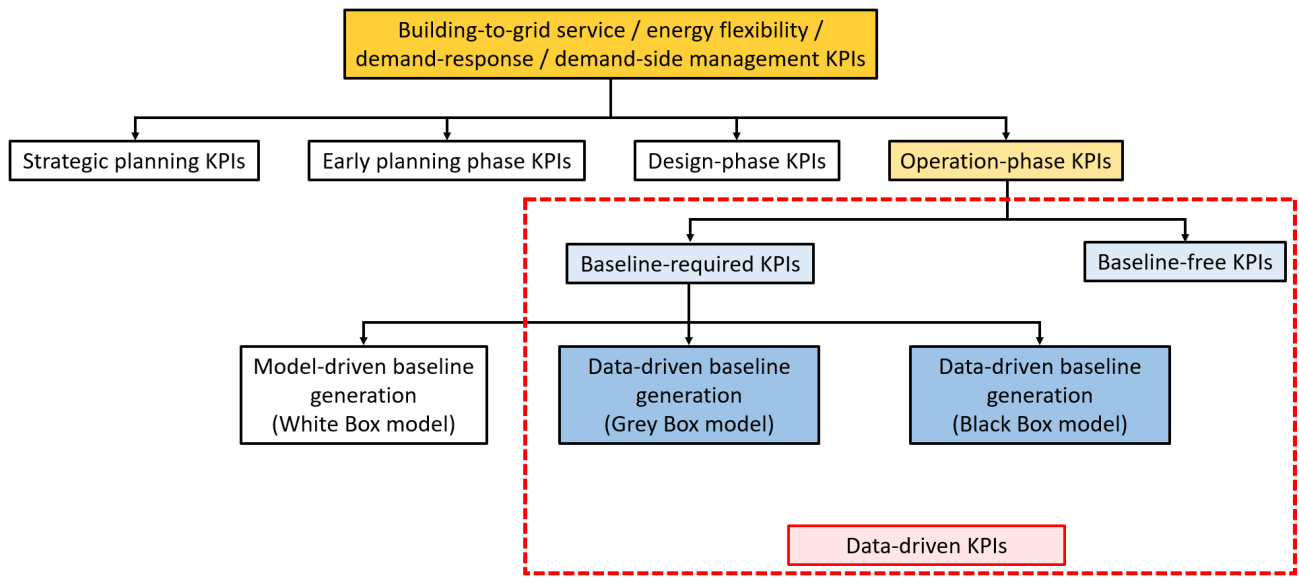

*Figure 1:* Scope of the KPI review.

*2.2. Literature search*

The review process in the current study mainly used the Web of Science engine to search for relevant publications. This was supplemented by queries with Google Scholar and Scopus, together with additional documents provided directly by the different authors of this article. *Table 1* shows the query sets used during the search. Note that the asterisk symbol (*) was used after some prefixes to allow searches for variations of a keyword. For example, "flexib*" enables search for both "flexible" and "flexibility" as keywords. The final query, which combined the query sets with the "AND" logic operator, was then conducted.

*Table 1:* Literature search query set.

| Query Set | Meaning |
|---|---|
| Topic = (building OR district OR community OR grid) | Building scope |
| Topic = ("energy flexib*" OR "demand flexib*" OR "demand-side management" OR "load shifting" OR "load shaving" OR "peak load reduction" OR "load modulat*") | Keywords related to energy flexibility definition |
| Topic = (KPI OR indicat* OR quantif* OR characteriz* OR metric) | Keywords related to energy flexibility quantification |
| Topic = (monitor* OR measure* OR "data-driven" OR "data driven") | Keywords related to data-driven approaches |

The query was conducted in March 2022. It identified 156 publications in total, with 1 technical report, 5 review articles, and 150 original journal papers. The initial 156 publications were then distributed among the different authors for deeper analysis. A total of 69 publications were discarded following the first-round review because they did not specifically investigate demand response or building energy flexibility topics. The remaining 87 papers were then gathered in a table which includes a brief summary of the use cases, applicable building sectors, flexibility resources, quantification methods, and potential stakeholders in each paper. This table with the complete list of the 87 selected reviewed articles can be found in *Table A1* of *Appendix A*. It serves as the foundation for the rest of the analysis carried out in this paper. The code/script used to analyze the data of *Table A1* can be found in *Appendix A*. *Figure 2* shows the number of publications categorized by year and corresponding scientific journals. The reviewed articles were published in 27 different journals and proceedings across different disciplines, including Applied Energy, Energy and Buildings, Energies, Energy and IEEE proceedings.

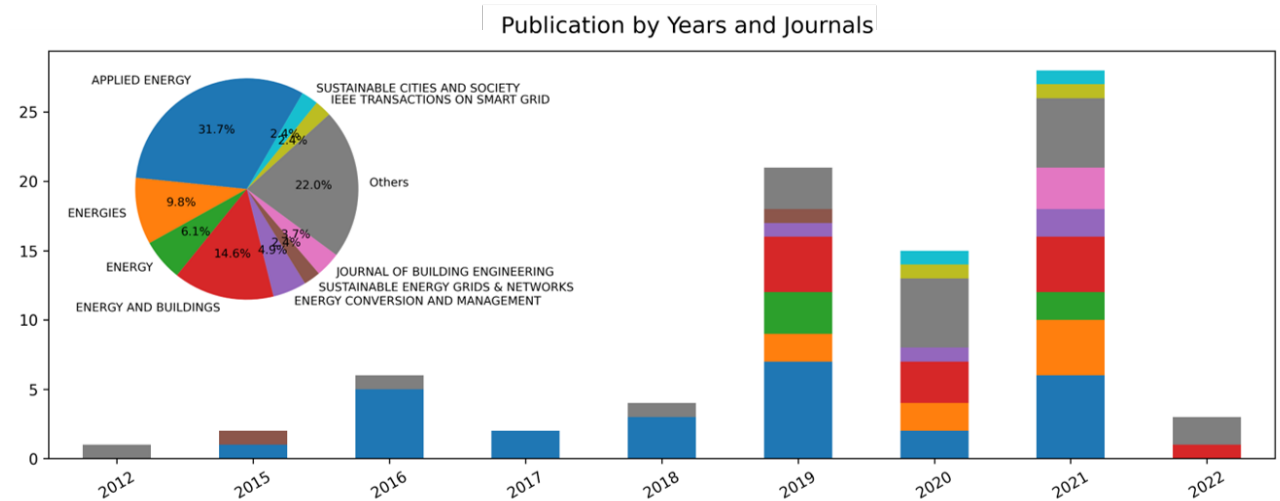

***Figure 2:*** *Distribution of the reviewed publications categorized by year of publication and type of scientific journal.*

From this *Table A1*, 48 distinct data-driven energy flexibility KPIs were identified and gathered and categorized in *Table A2* (see *Appendix A*) according to their equation and definition similarities. In addition, 29 other generic building KPIs associated with energy flexibility studies in the reviewed publications were collected and included in *Table A2*. *Table A2* includes the definition, formula, and relevance of the KPIs, together with an indication of the performance aspects, type of flexibility, needed input variables, computation complexity, and popularity among existing studies. The KPI analysis in this review was based on the information collected in *Table A1* and *Table A2*.

The study and categorization of the different publications and KPIs in this article are based on the collegial analysis and discussions of the authors. All authors of the present study have strong expertise in the field of building energy flexibility and demand response. They also are active members of the building energy flexibility community and participate in related international research projects such as the IEA EBC Annex 67, 81, 82 and 84.

### 3. Use cases for data-driven energy flexibility KPIs

A systematic review was carried out to determine key characteristics, target stakeholders, and types of technologies present in the different studies using data-driven energy flexibility KPIs. These publications can be analyzed according to three main axes:

- **Building sector:** Commercial, residential, or industrial buildings

- **Scope:** Two main levels: single building or cluster of buildings

- **Assessment method:** Simulation, measurement, or hardware-in-the-loop (HIL)

As observed in *Figure 3*, the sector of residential buildings was the most studied (48.9%), followed by the commercial sector (28.3%). One can note that although modern building management systems (BMS) collecting building performance data are more common in commercial buildings than in residential ones, the former are less studied than the latter. This can be because energy flexibility operation strategies are easier to implement in residential buildings and have tremendous potential for demand-side management [21].

In addition, single-building level application studies (53.3%) are more common than cluster-level ones (41.3%). This can be explained by the complexity of setting research projects with clusters of buildings before having results on single buildings to motivate the realization of the former. Small-scale demand response investigations on single buildings are easier, and thus more common at the moment. However, based on the promising results obtained on large-scale applications [22][23], an increased interest in cluster-level studies is expected in the near future.

The building operational data for energy flexibility assessment come from different sources. Only 25.8% involved real measurements, and 65.2% relied purely on simulations. The fact that simulation is still the most used investigation method indicates the need for studies with real case study applications.

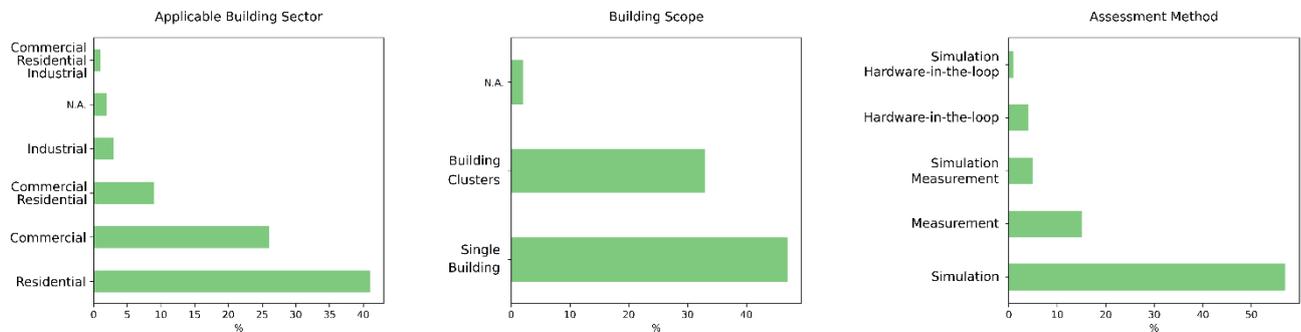

*Figure 3: Distribution of the reviewed studies using data-driven energy flexibility KPIs according to the building sector, building scope, and assessment method.*

According to the literature review, energy flexibility KPIs are typically used for the assessment of the following technical solutions: the control of heat pumps [24][25][26][27], district heating [28], HVAC systems [29][30][31], charging and discharging of EVs and batteries [32][33][34], and lighting [35]. It was also possible to distinguish a focus on the design phase and operational phase. It was found in previous reviews [7] that for the design phase, the influence of the building envelope [36][37][38][39] on energy flexibility potential was the most popular topic. For the operational phase, the coordination between different technologies [24][27] and the indoor zone temperature setpoint adjustments [36][38][40][41] were the most common management strategies found to exploit building energy flexibility.

Moreover, the stakeholders interested in the flexibility at the single-building level are the occupants, building owners, and building managers, while DSO/TSO and utility companies focus on the building cluster level. In both levels of application, the results represented in the flexibility indicators are crucial for policymakers and planners. In this regard, metrics and methodologies on how to evaluate energy flexibility are becoming key factors in improving energy management at both the grid and user levels. Knowledge of the impact of each building on the grid can provide insights into where to act to optimize the operation of the whole energy system.

### 4. Data-driven energy flexibility KPIs

The current review focuses on 87 papers obtained from the initial screening (see *Table A1*). A total of 81 data-driven KPIs were extracted from those publications (reduced down to 48 distinct KPIs after the merger of the ones with a similar definition, equation or underlying logic but different form), which cover a wide spectrum of applications with various building cases, performance goals, and data requirements. While some KPIs are more popular and commonly used across different studies, others were developed specifically for unique scenarios. To structure and clearly understand the landscape of data-driven building energy flexibility assessments elaborated by the scientific community, these 48 KPIs were systematically categorized based on the following criteria:

- **Relevance:** Some KPIs are designed specifically for assessing energy flexibility, while others are more generic. The KPIs were qualitatively rated as "low," "medium," and "high" based on how relevant they are to the best of the authors' knowledge.

- **Complexity:** The computation of KPIs involves data collection, processing, and calculations. The complexity of the KPIs was qualitatively assessed based on the required amount of data processing and computation.

- **Performance aspects:** Depending on the use cases, KPIs may have different primary performance aspects. The latter are categorized into energy demand, power demand, cost, greenhouse gas (GHG) emissions, impacts on IEQ, and comfort.

- **U.S. DOE categorization:** The U.S. DOE categorized DSM strategies into five types: efficiency, shifting, shedding, modulation, and generation. Each KPI was tagged with the relevant U.S. DOE categorization.

- **Baseline:** Each KPI was marked as either baseline-required or baseline-free, according to whether a baseline scenario is needed for its computation or not (as introduced in Section 2).

- **Data requirements:** Depending on the application scopes and use cases, energy flexibility KPIs have diverse input data requirements for their computation. The temporal and spatial data requirements, as well as the variable types needed for the KPIs' calculation, were also investigated.

The categorization resulted in 12 core energy flexibility categories with 48 data-driven KPIs and four generic categories with 29 KPIs. One should note that core KPIs are specifically designed for energy flexibility assessments, while generic KPIs are not directly linked to energy flexibility assessment but are often used in the reviewed studies to evaluate other performance aspects of buildings involved in DR. All KPIs with detailed categorization can be found in *Table A2*. The analyses in the following sections focus on these 48 core energy flexibility KPIs. The popularity of these KPIs has also been evaluated by counting the number of publications using them. A summary of the classification of these KPIs can be found in *Table 2*.

***Table 2:*** *Summary of the categorization for both the energy flexibility KPIs (EF KPI) and the associated generic building performance KPIs (Generic KPI). These KPIs have been extracted from the 87 selected papers, then categorized. The information of this table is related to distinct KPIs: KPIs with similar definitions, equations, or underlying logic but different forms were merged.*

| Category | Number of distinct KPIs | Number of baseline-required KPIs | Number of baseline-free KPIs | Popularity: number of publications |
|---|---|---|---|---|
| EF KPI: Peak power shedding | 4 | 4 | 0 | 8 |
| EF KPI: Energy / average power load shedding | 10 | 8 | 2 | 11 |
| EF KPI: Peak power / energy rebound | 3 | 3 | 0 | 5 |
| EF KPI: Valley filling | 2 | 2 | 0 | 1 |
| EF KPI: Load shifting | 6 | 2 | 4 | 14 |
| EF KPI: Demand profile reshaping | 3 | 2 | 1 | 4 |
| EF KPI: Energy storage capability | 3 | 2 | 1 | 11 |
| EF KPI: DR energy efficiency | 4 | 4 | 0 | 7 |
| EF KPI: DR costs / savings | 4 | 4 | 0 | 10 |
| EF KPI: DR emission / environmental impact | 2 | 2 | 0 | 3 |
| EF KPI: Grid interaction | 3 | 2 | 1 | 3 |
| EK KPI : Impact on IEQ | 4 | 4 | 0 | 5 |
| Generic KPI: Energy efficiency | 6 | 2 | 4 | 3 |
| Generic KPI: Costs and savings | 4 | 1 | 3 | 7 |
| Generic KPI: $CO_2$ emissions / environmental impact | 2 | 2 | 0 | 3 |
| Generic KPI: Grid interaction | 17 | 6 | 11 | 13 |

*4.1. Baseline-required KPIs*

*4.1.1. Baseline-required data-driven KPIs*

Out of the 48 collected and categorized data-driven energy flexibility KPIs (see *Table 2*), 39 of them (81%) were found to be baseline-required. *Figure 4* shows the distribution of those baseline-required KPIs according to both the U.S. DOE categorization of DSM and the relevance assessment by the authors of the present review. The y-axis indicates the number of KPIs and their relevance in each category. It can be observed that load shedding and load shifting are the most popular categories, covering 52% of the KPIs. Some of those KPIs were specifically developed for a certain type of assessment: 10 are designed for load shedding only, 4 for load shifting only, and 2 for energy efficiency only. For instance, the Flexibility Index [38] was specifically created to measure the capacity of a system to shift energy use from high-price periods to low-price periods. On the contrary, more than half (23 out of 39) of the baseline-required KPIs are for a more general assessment of energy flexibility and cover multiple categories. For example, the Flexible Savings Index [42] quantifies the differences in cost savings between penalty-aware operations (flexible) and penalty-ignorant operations (baseline). Since it is not restricted to operational cost reduction, this KPI can be used to assess energy efficiency, load shedding, and load shifting performance during a DR event.

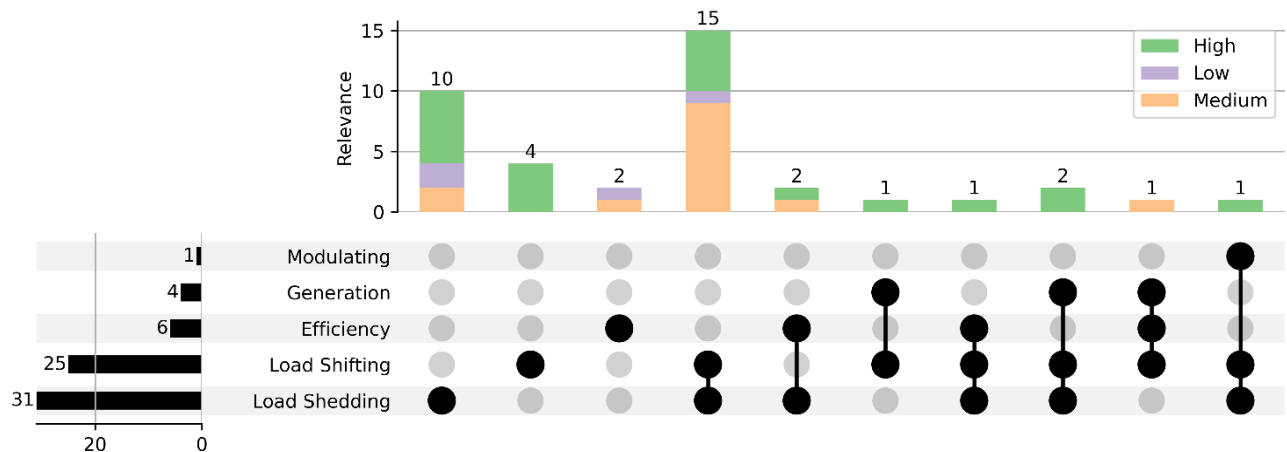

***Figure 4:*** *Baseline-required energy flexibility KPIs analyzed according to the U.S. DOE categorization of DSM and the relevance assessment from the authors of the present review.*

*Figure 5* shows the distribution of complexity, temporal evaluation windows, and spatial resolution of the baseline-required KPIs. This figure gives an overview of some of the key characteristics of baseline-required KPIs and can thus inform on the common scope and applicability of the latter. 82% of these baseline-required KPIs have low complexity, meaning they are easy to calculate with baseline and flexible performance data. As for the temporal evaluation window, 56% of the KPIs can be applied to a single DR event only, while the rest are less restricted in terms of the temporal aspects. 15% of them assess energy flexibility for a whole day and 8% for a whole year. Furthermore, 5% of the KPIs can be used for multiple DR events, and 15% have no duration assessment restriction. In terms of applicable spatial resolution, about half of the KPIs are for single buildings (31%) and building clusters (21%), while the rest are not restricted to any spatial scale.

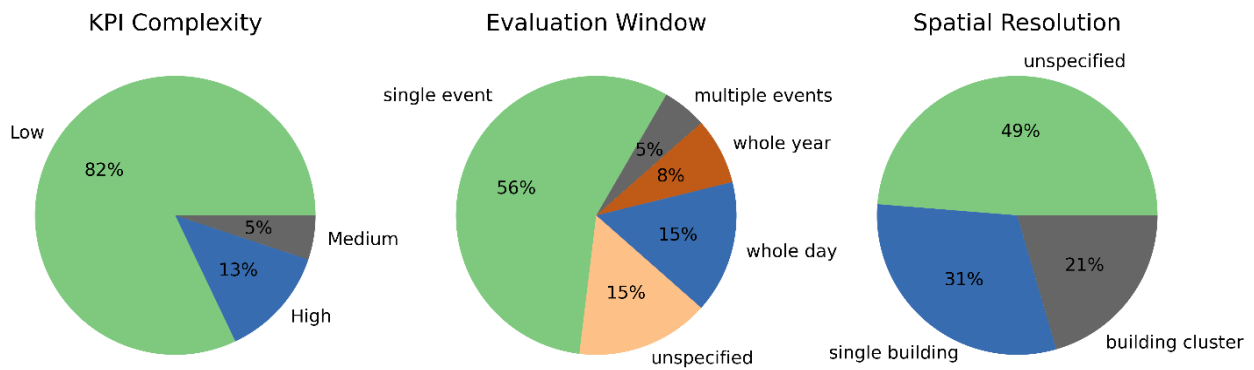

***Figure 5:*** *Distribution of the baseline-required energy flexibility KPIs according to their complexity, temporal evaluation window, and spatial resolution.*

*4.1.2. Data-driven methods to generate baselines*

Out of the 48 KPIs identified previously, 39 of them (81%) required the generation of a baseline for energy flexibility quantification [43]. The baseline is an estimation of the consumers' "normal" energy usage, i.e., the energy demand profile without any DR event [44]. A good baseline estimation methodology should be robust and transparent and with an acceptable level of accuracy [45]. It also should limit the opportunity for "cheating" and market manipulation in case of reward based on the baseline estimation [45].

This section reviews data-driven baseline generation methodologies applicable to single-building and district levels (aggregated energy demand). This review is not meant to be exhaustive on the topic of baseline estimation but provides insights and key references for methodologies that can be used for flexibility characterizations that are based on a comparison against a baseline. Baselining is closely related to the topic of short-term load forecasting. The main difference is the possibility of including post-DR event measurements in the estimation. According to recent publications on the topic, the methodologies used for baseline estimation can be classified as: control group, averaging, regression, machine learning, and hybrid [45][46][47][48].

The control group methods require monitoring a similar group of buildings, for which no flexibility scenario is applied. The monitoring data of this control group is then used as a baseline for the characterization of other buildings performing DR. The averaging method (also called *similar day look-up approach* or *XofY*) is one of the oldest forecasting methods and is still widely used for baseline estimation [49]. The most popular methods are the High3of5 and Mid5of10. HighXofY takes the average load of the X highest consumption days from a set of Y admissible days preceding the DR event. The exponential moving average (or exponential smoothing) method is a weighted average of the customer's historical load, where the weight decreases exponentially over time [49].

Regression models, either linear or higher order nonlinear, are often used to estimate customer baseline load due to their robustness but require a relatively large historical dataset to be fitted correctly. The feature selection of regression models is crucial and varies significantly in the literature [50][51]. The most common features encountered are the following: historical load variables, external variables (weather- and time-related), and building and occupancy characteristics. Autoregressive models (ARMA, ARIMA) are special forms of regression models that are widely used for load forecasting. They evaluate the current value of the series as a linear combination of previous/past loads. Such models can handle seasonality and non-stationarity and only require a limited amount of historical data [52]. Other types of regression models exist, such as the GAM (Generalized Additive Model) and LASSO (Least Absolute Shrinkage and Selection Operator) types. These have been used for energy flexibility KPI computation [53].

More recently, a variety of shallow learning and deep learning-based methods have been employed for energy profile forecasting. In particular, support vector machines for regression (SVR), extreme gradient boost (XGBoost) and random forest are popular shallow learning methods for building load predictions. Vanilla deep neural networks (DNN), long short-term memory (LSTM) and time-delay neural networks (TDNN) are common deep learning methods for load forecasting [54]. Regarding the aforementioned DNN methods, however, Antonopoulos et al. [54] comment on the need for a large amount of training data to outperform other more robust statistical methods.

Hybrid models also have been developed to combine different forecasting approaches. In Denmark, for instance, a flexibility response model was trained with field data from a portfolio of 138 real residential customers equipped with electric heaters. This model was based on a linear combination of three methods: linear interpolation, forward-backwards autoregression, and load decomposition [45].

A comparison between methods is not straightforward, as there is not a single way to define/set each of them, and their respective input data and parameters can be quite different. Many publications only use a single baselining method, which makes the comparison difficult. Moreover, the case study or the metrics used to evaluate the performance of a model are not homogenous among articles. However, some conclusions can be drawn from the literature. It is difficult to disregard any of the methodologies described above [55]. As highlighted by Makridakis and Hibon [56], "simple methods developed by practising forecasters do as well, or in many cases better, than sophisticated ones." Some articles conclude that the simple baseline methodologies provide a good and simple basis for developing customer baseline load [49]. Many industrial solutions for demand response rely on simple baseline methodologies to reward customers, such as High3of5 or Mid5of10. Indeed, they state that the simplicity and transparency of these types of methodologies make them reliable and understandable by customers [50][57]. However, more field studies are required to evaluate the most appropriate baselining methodologies.

In addition, the scale of aggregation plays an important role in model selection. The prediction can be quite poor at the household level due to the high variance and stochasticity of occupants (errors ranging from 5% to 60% [58]). As stated by Humeau et al. [51], SVR might be the most suitable method for load forecasting of a small district (782 houses), but linear regression performs better at the household level. Peng et al. [59] conclude that "the selection of load forecasting techniques highly depends on the data itself, and there is no single technique that outperforms other techniques in all scenarios, especially for load at low aggregation levels."

*4.2. Baseline-free data-driven KPIs*

Unlike baseline-required KPIs, baseline-free KPIs can be calculated with building data measured in a single scenario. Only nine data-driven baseline-free energy flexibility KPIs were identified from the review. *Figure 6* shows their distribution according to the U.S. DOE categorization of DSM and the relevance assessment from the authors of the present review. Specifically, six of them (67%) have at least medium relevance, which is a lower percentage than the baseline-required KPIs (89%). Among them, except for the Available Flexible Energy [24], which covers both load shifting and load shedding, the rest of the KPIs are dedicated to load shifting [36][60][61][62][63], load shedding [28][32][45][64][65][66][67][68], and modulating [22][67], respectively. There is no baseline-free KPI for efficiency and generation.

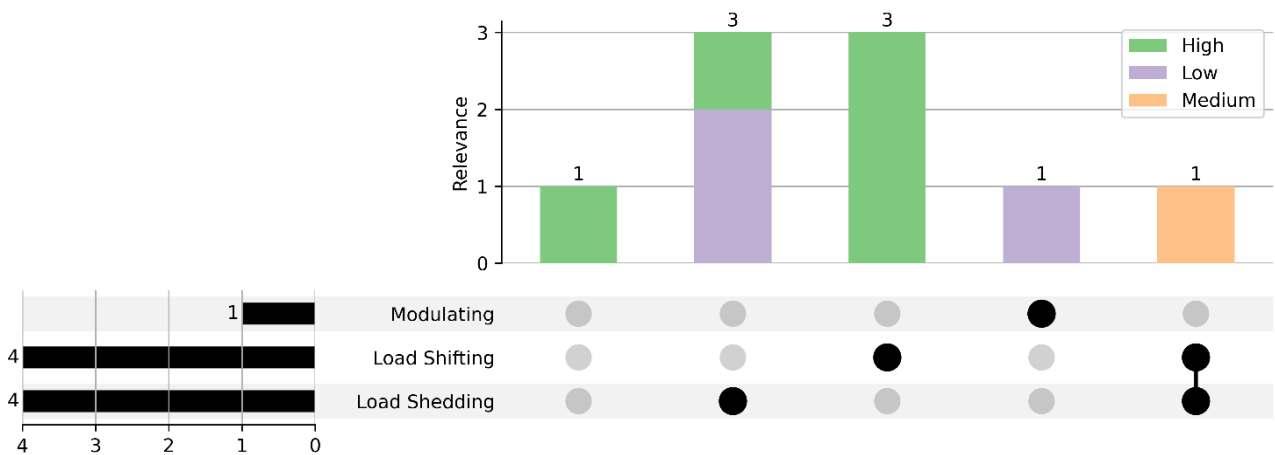

***Figure 6:*** *Baseline-free energy flexibility KPIs analyzed according to the U.S. DOE categorization of DSM and the relevance assessment from the authors of the present review.*

*Figure 7* presents other statistics of the baseline-free KPIs. A smaller portion of the baseline-free KPIs (56%) have low complexity compared with baseline-required KPIs (82%). This is because those KPIs are more likely to involve sophisticated data manipulations. However, one should not ignore the amount of effort needed in developing data-driven models for the baseline-required KPIs. In terms of the temporal evaluation window, four KPIs are intended to a single DR event, two KPIs are only applicable to yearly-level evaluations, and three KPIs can be used for an arbitrary event window. As for the spatial resolution, except for one KPI that was designed for a single building, the rest of the eight KPIs are not restricted to a specific number of buildings.

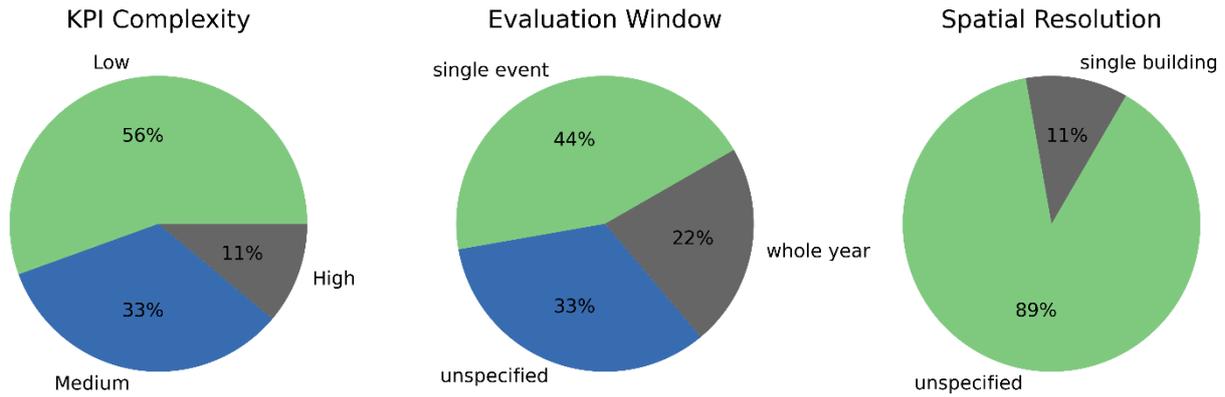

***Figure 7:*** *Distribution of the baseline-free energy flexibility KPIs according to their complexity, temporal evaluation window, and spatial resolution.*

Baseline-free KPIs are especially useful when it is difficult to monitor the performance of a building during what is supposed to be a baseline scenario or when no data-driven approach can be used to generate that baseline. Indeed, buildings enabling energy flexibility strategies might perform DR very frequently or continuously. In that case, it might be impossible to record the performance of that building without any DR event over a sufficient period of time and with adequate boundary conditions to be used as a baseline for assessing DR events. A popular baseline-free KPI is the Flexibility Factor (FF), which was first defined by Le Dréau & Heiselberg in 2016 [36], with multiple subsequent variations [60][62][63]. The FF assesses how well an operational strategy could shift a target quantity outside of a temporal window. The FF can be applied in different cases because the target quantity can be energy usage, operational costs, carbon emissions, and even HVAC system runtime. However, as stated above, most existing studies rely on simulations to get both baseline and DR scenarios. Future efforts are still needed to develop KPIs that are baseline-free, easy to calculate, and highly relevant.

*4.3. How to choose data-driven KPIs?*

*Figure 8* shows the identified data-driven energy flexibility KPIs and their relation to the different stakeholders. One can observe that the majority of the KPIs are intended for energy grid operators.

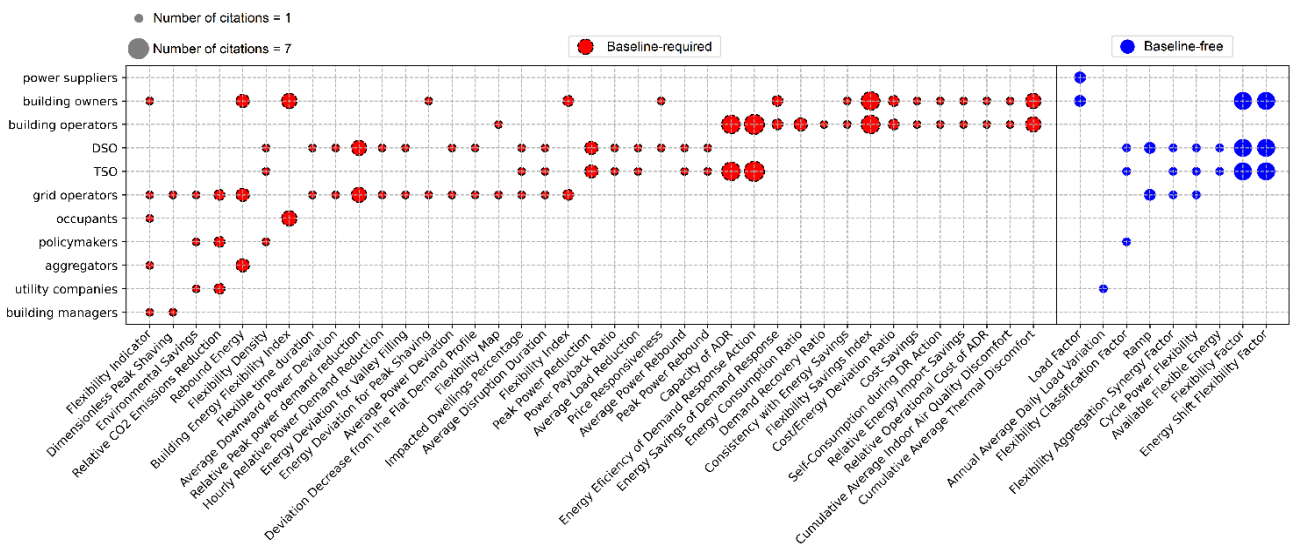

***Figure 8:*** *Data-driven energy flexibility KPIs analyzed with regard to the different stakeholders. The popularity of the different KPIs was assessed by the number of publications/citations in which they appeared.*

Furthermore, those KPIs have diverse applications across different scopes and have various data requirements. The parallel categorical plot in *Figure 9* presents the multidimensional categorization of the 48 main data-driven energy flexibility KPIs. One can understand a KPI's categorization by following the line connecting the columns. For example, the Flexibility Savings Index (FSI) evaluates the cost savings of a DR event and was used in six publications. The FSI is highly relevant and with a low calculation complexity. However, a baseline scenario is required to compute it. In contrast to some KPIs that are designed for a single DR event, the FSI can be evaluated for an arbitrary timespan and can be calculated from building data time series with various levels of granularity. The FSI can be used for both single buildings and building clusters.

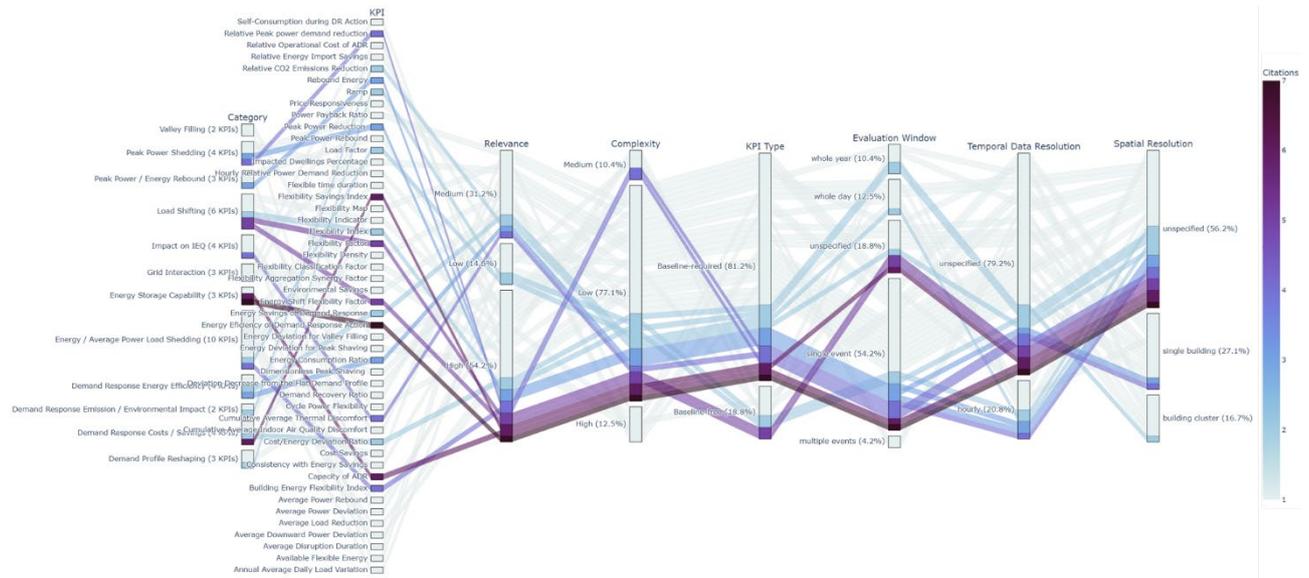

***Figure 9:*** *Categorization of the data-driven energy flexibility KPIs. The popularity of the different KPIs is assessed by the number of publications/citations in which they appear.*

The analyses in the sections above clearly indicate the large diversity of the data-driven energy flexibility KPIs. It can thus be quite challenging to choose an appropriate one for a given situation. The authors of the present review, therefore, suggest following the few steps below to select appropriate KPIs (summarized in *Figure 10*):

1) Identify the targeted stakeholders: e.g., TSO, DSO, building operators or building occupants.
2) Determine the application scope: single building or building cluster.
3) Determine the main goal of the energy flexibility measures: e.g., reducing peak power demand or load shifting to reduce operational cost.
4) Check if baseline-free KPIs are sufficient: those KPIs are usually much easier to calculate because they do not require extra effort for the generation of a baseline scenario.
5) If baseline-free KPIs are insufficient, a data-driven modelling approach needs to be employed to generate the baseline scenario of the target building before the KPI can be computed.
6) If the existing data-driven KPIs are deemed inappropriate for the current needs, a new tailored KPI should be developed.

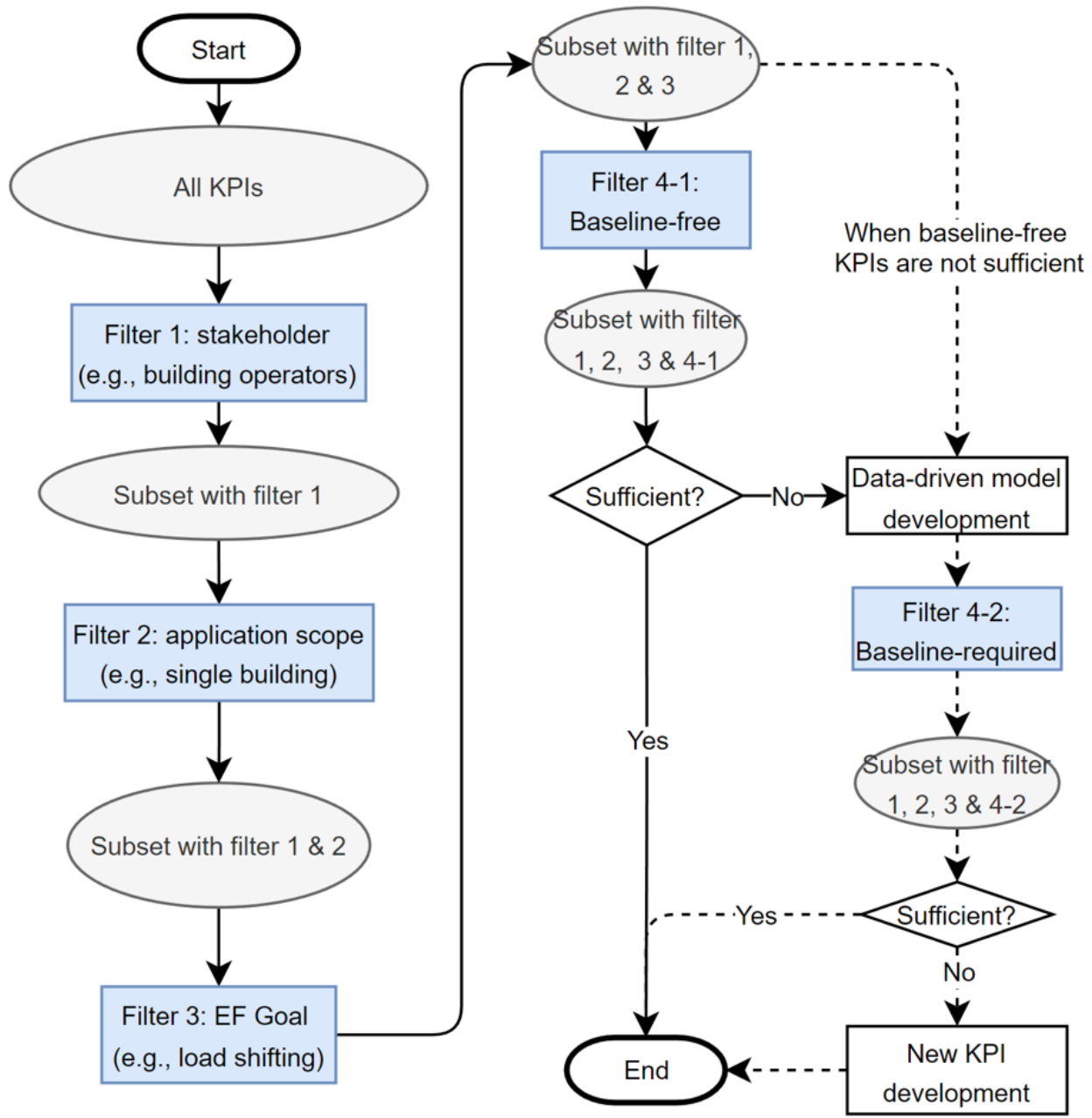

***Figure 10:*** *Selection process of data-driven energy flexibility KPIs.*

One can find in *Table 3* the top three most popular baseline-required and baseline-free data-driven energy flexibility KPIs. One should note that the popularity of KPIs in the scientific literature is not a guarantee of their adequacy and usefulness. However, until systematic analyses of these KPIs are performed with various datasets of building performing DR, this table can serve as a guideline to get an overview of which data-driven energy flexibility KPIs are commonly used by the scientific community.

*Table 3: Top 3 baseline-required and baseline-free data-driven energy flexibility KPIs.*

| | KPI | Formula | Terms | Stakeholders | Performance goals | U.S. DOE categorization [16] | Reference |
|---|---|---|---|---|---|---|---|
| **Baseline-required** | Energy Efficiency of Demand Response Action | $\eta_{ADR} = 1 - \frac{\int_0^\infty (Q_{ADR} - Q_{ref})dt}{\int_0^{length_{ADR}} (Q_{ADR} - Q_{ref})dt}$ | $Q_{ADR}$: thermal power supplied to the building during the DR event; $Q_{ref}$: thermal power supplied to the building during reference operation | Building operators; TSO | Reduce energy consumption during the DR event | Energy efficiency | [34][41][68][69][70][71] |
| | Flexibility Savings Index | $FSI = \frac{Cost\ of\ flexible\ operation}{Cost\ of\ baseline\ operation}$ | - | Building owners; building operators | Reduce operational costs during the DR event | Energy efficiency, load shedding, load shifting | [29][42][72][73][74][75] |
| | Peak Power Shedding | $\Delta P = P_{baseline,peak} - P_{flexible,peak}$ | $P_{baseline,peak}$: the peak power demand during the peak hour of the baseline scenario; $P_{flexible,peak}$: the peak power demand during the peak hour of the flexible scenario | DSO; TSO; grid operators | Reduce power demand during peak hour due to flexible operation | Load shedding | [68][76][77][78] |
| **Baseline-free** | Flexibility Factor | $FF = \frac{\int q_{non\ peak} \cdot dt - \int q_{peak} \cdot dt}{\int q_{non\ peak} \cdot dt - \int q_{peak} \cdot dt}$ | $q_{non\ peak}$: the quantity of interest during non-peak periods; $q_{peak}$: the quantity of interest during peak periods. | DSO; TSO; building owners | Shift a quantity between periods | Load shifting | [36][62][63][72][75][79] |
| | Energy Shift Flexibility Factor | $FS = \frac{\int q_{heating\ (low\ price)} \cdot dt - \int q_{heating\ (high\ price)} \cdot dt}{\int q_{heating\ (low\ price)} \cdot dt + \int q_{heating\ (high\ price)} \cdot dt}$ | $q_{heating\ (low\ price)}$: the heating demand during low-price periods; $q_{heating\ (high\ price)}$: the heating demand during high-price periods | DSO; TSO; building owners | Shift energy consumption between periods | Load shifting | [35][36][60][61][80] |
| | Load Factor | $LF = \frac{AVG_L}{max_L}$ | $AVG_L$: the average demand during a period; $max_L$: the maximum demand during a period | Building owners; power suppliers | Reduce the peak power demand | Load shedding | [73][81] |

The 29 KPIs categorized as "generic" were not initially designed for assessing building energy flexibility but are often used in studies investigating energy flexibility and Grid-interactive Efficient Building (GEB). Two commonly found examples of these "generic" KPIs are self-consumption and self-sufficiency. Although originally developed to evaluate on-site renewable energy generation and utilization, many studies [82][83][84] used them to indicate the flexibility of a building's energy use in response to available on-site production.

## 5. Available datasets for testing and analyzing energy flexibility KPIs

This section curates a list of surveyed datasets performing demand response or building-to-grid (B2G) services for energy flexibility assessments. The proposed process includes preparing the collected datasets with proper descriptions and characterizing them in terms of their DSM strategy, building scope, grid type, and control actions and objectives. This section also assesses their capability to calculate the energy flexibility KPIs by referring to their available features compared to the ones each KPI requires. This is a significant step towards testing and benchmarking the energy flexibility KPIs reviewed above. To the best of the authors' knowledge, this survey and analysis is the first standardized effort to curate open datasets for quantifying building energy flexibility. Although limited in size, this gives a good overview of the features commonly found in DR-related projects publishing data. The authors also hope that this first collection effort can seed and improve the generation, description and publication of future open-access datasets of buildings performing DR.

*5.1. Dataset collection process*

This section analyzes datasets collected from research studies and pilot projects on B2G application services. The scope for this dataset collection included data (in the form of time series, along with the appropriate metadata and case description) from existing, simulated, or semi-simulated (hardware-in-the-loop) individual buildings, multiple buildings or building clusters that participate in demand response, demand-side management, or energy flexibility studies or programmes. The collection process was based on: (1) a search of publicly available datasets and data platforms (e.g., Kaggle, Data-in-Brief); (2) open calls published on online platforms and social networks and shared with the IEA EBC Annex 81, 82, 83, and 84; and (3) personal contacts to lead investigators of research studies being considered within IEA EBC Annex 81 and 82 activities. The registration of the datasets was obtained through a semi-structured online questionnaire, which proposed standardization of nomenclature and terms to characterize the description of the datasets (i.e., metadata and case information/characteristics).

A breakdown of the dataset collection process and quality assessment is presented via a Sankey diagram in *Figure 11*. A total of 330 datasets were identified. Half of the considered datasets were deemed out of scope, mostly because they did not comprise buildings actually performing DR, despite some initial promising descriptions of the studies. Of those remaining, only 20% (33) of the project teams that were contacted (to ascertain if they could provide information and data from their studies) responded. Approximately 30% (10) of the respondents were unable to share datasets due to confidentiality agreements, and 21% (7) of the others had inadequate datasets. Of the remaining, 63% (10) were able to provide a partial or full description of their datasets but did not share or publish the datasets at the time of writing. Only 37% (6) were able to provide access to the dataset along with a partial or full description of the dataset.

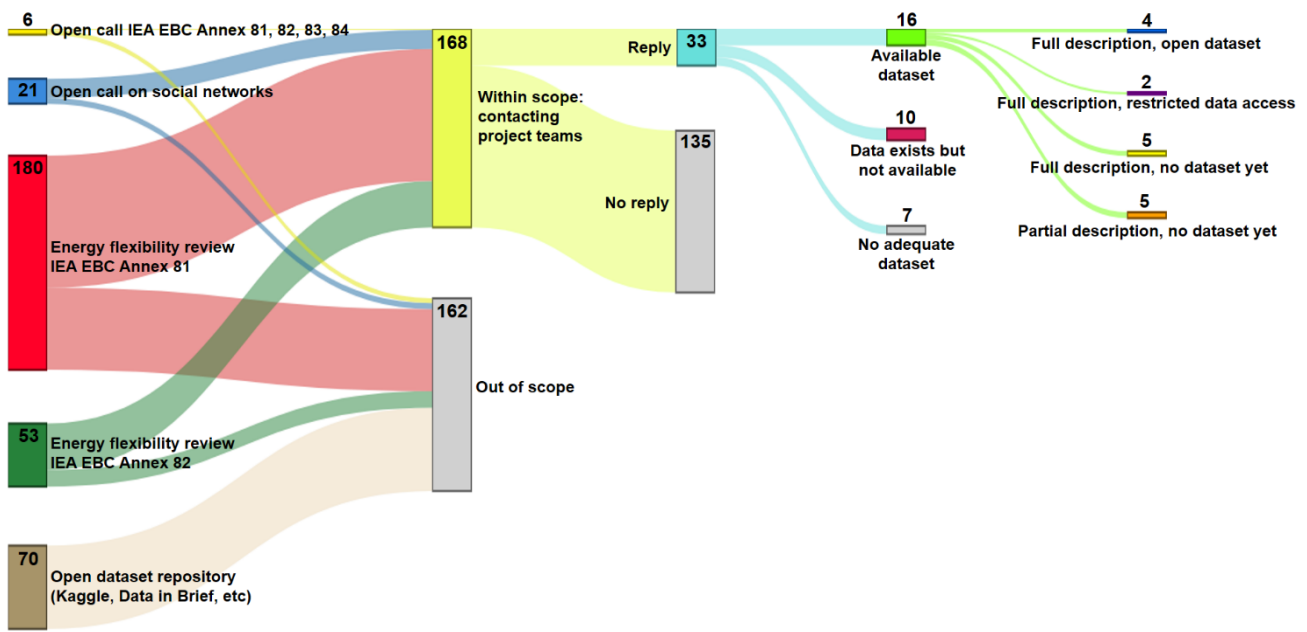

*Figure 11: Breakdown of the datasets and information collection process and quality assessment.*

The analysis presented in the following sections only concerns the 16 (4.8% of the in-scope total) datasets, where the associated research teams provided or will provide the necessary access to the appropriate data. The list and detailed information about these 16 datasets can be found in *Appendix B, Table B1*. The code/script used to analyze the data of *Table B1* can be found in *Appendix B*. A summary of the information on the 16 datasets of buildings performing DR can be found in *Table 4*.

*Table 4: Information summary of the 16 datasets of buildings performing DR.*

| Status | Project name | Institution | Country | Case description | Case type | Link to dataset | Reference |
|---|---|---|---|---|---|---|---|
| Full description, dataset available | A medium-sized office building in Berkeley | LBNL | USA | 1 low-rise office building | Real monitored building | https://datadryad.org/stash/dataset/doi:10.7941/D1N33Q | [18] |
| Full description, dataset available | Domestic hot water usage data | Stellenbosch University | South Africa | 77 residential houses | Real monitored building | https://bit.ly/synthetichotwater | [85] |
| Full description, dataset available | La Rochelle Residential District | La Rochelle University | France | 337 dwellings in 98 large apartment blocks | Simulation | https://gitlab.univ-lr.fr/jledreau/AtlanFlex-Smart-thermostat | [86] |
| Full description, dataset available | INCITE | IREC | Spain | 1 large multi-family apartment block | Hardware-in-the-loop | https://doi.org/10.5281/zenodo.7006826 | [87] |
| Full description, dataset available with restricted conditions | SEIH | Aalborg University | Denmark | 191 detached single-family houses | Real monitored building | - | [88] |
| Full description, dataset available with restricted conditions | Aliunid | EMPA | Switzerland | 1 residential building | Real monitored building | - | [89] |
| Full description, dataset not shared yet | Newcastle Energy Center | CSIRO | Australia | 1 low-rise office building | Real monitored building | - | - |
| Full description, dataset not shared yet | Simulated Building to Distributed Network | Syracuse University | USA | 400+ residential and commercial buildings | Simulation | - | [90] |
| Full description, dataset not shared yet | Energy flexibility Danish building stock | Aalborg University | Denmark | 20+ residential buildings | Simulation | - | - |
| Full description, dataset not shared yet | Hermandades neighborhood retrofit | Universidad nacional de La Plata | Spain | 102 low-rise apartment blocks, 24 high-rise apartment blocks, 2 | Simulation | - | [78] |

| | | | | schools, 1 sport field | | | |
|---|---|---|---|---|---|---|---|
| Full description, dataset not shared yet | CityLearn-RBC/MARLISA Simulation | University of Texas at Austin | USA | 4 commercial buildings, 5 apartment blocks | Simulation | - | [91][92] |
| Partial description, dataset not shared yet | Institutional nZEB in Montreal doing MPC | Varennes Library | Canada | 1 net-zero public library building | Real monitored building | - | - |
| Partial description, dataset not shared yet | Atlanta, HIL | Drexel University, Texas A&M University, NIST | USA | 1 low-rise office, 1 mid-rise office | Hardware-in-the-loop | - | - |
| Partial description, dataset not shared yet | Bufalo, HIL | Drexel University, Texas A&M University, NIST | USA | 1 low-rise office, 1 mid-rise office | Hardware-in-the-loop | - | - |
| Partial description, dataset not shared yet | New York, HIL | Drexel University, Texas A&M University, NIST | USA | 1 low-rise office, 1 mid-rise office | Hardware-in-the-loop | - | - |
| Partial description, dataset not shared yet | Tucson, HIL | Drexel University, Texas A&M University, NIST | USA | 1 low-rise office, 1 mid-rise office | Hardware-in-the-loop | - | - |

*5.2. Datasets description and analysis*

The 16 collected datasets represent a wide variety of B2G studies. The data are characterized by low availability, heterogeneous formats and diverse end-use domains. They include data from real monitored buildings (6), hardware-in-the-loop setups (5), and numerical simulations (5). The buildings are located in the USA (7), Europe (6), Canada (1), Australia (1), and South Africa (1). All building typologies are represented, including non-residential buildings (7), residential buildings (6), and clusters combining both types (3). Five datasets comprise a single building, five datasets comprise 1–10 buildings, two datasets comprise 10–100 buildings, and four datasets comprise 100–500 buildings. While most datasets are associated with electrical grids (15), only a few (2) are connected to district heating networks. Regarding the tariff programs/structures, most cases are based on time-of-use (7), real-time pricing (6), and flat-rating pricing (2), with four lacking this information. Some of the datasets are based on a time duration of one to four years (6) and others of one to four months (6); two datasets were accepted exceptionally, with only a few days of data. The remaining datasets did not specify this information. The temporal resolution (sampling rate) of the datasets ranged from one minute (7), to sub-hourly (4), and hourly (3); two datasets did not specify. The majority of the datasets had less than 1% of the data points missing. Some did not have the necessary information to make this assessment.

The most common features available across datasets are indoor and outdoor temperatures, followed by end-use energy demand, thermostat setpoints, occupancy, and solar radiation. Approximately 40% of the datasets included data related to heating loads and cooling loads (i.e., chilled water temperature). Another standard available data feature is the price signal used for activating most of the flexibility control strategies (load shifting, load shedding, generation, and modulation) and assessing rebound effects and the impact on flexibility during and after the DR events. Regarding additional grid signals, some datasets include the duration of the event. However, none provide the requested capacity or financial incentives typically included in bilateral transactions or agreements between energy market stakeholders.

*Figure 12* illustrates the usability of the reviewed KPIs (the share of collected datasets that can be used to compute the different KPIs) and the usefulness of the datasets (the share of KPIs in each category that can be computed from the data in each dataset). The five most easily calculated KPIs in terms of required and available variables are energy/average power load shedding, load shifting, demand profile reshaping, demand response energy efficiency, and demand response costs/savings. The most commonly used variables when calculating the KPIs include end-use energy demand, power demand, price signal, event request time period (interval), action type/direction (downward and upward flexibility), and size (flexibility capacity). While the last three are critical parameters, most datasets do not include more than one of these three. As a result, the majority of KPIs cannot be calculated directly with most datasets (i.e., not without performing additional modelling and/or calculations to derive the required variables). It is interesting to note that the value of a dataset for KPI calculation does not increase with the number of features it contains. While datasets #1, #11, and #12 have more available features, *Figure 12* shows that datasets #2, #3 and #6 are the top three for calculating the most KPIs. This is illustrated in the upper and bottom subplots of *Figure 12*. For example, in the upper subplot, dataset #6 can be used to calculate 90% of the KPIs for energy/average power load shedding, 33% for load shifting, and so on. In the bottom subplot, dataset #6 can be used to compute 56% of all KPIs.

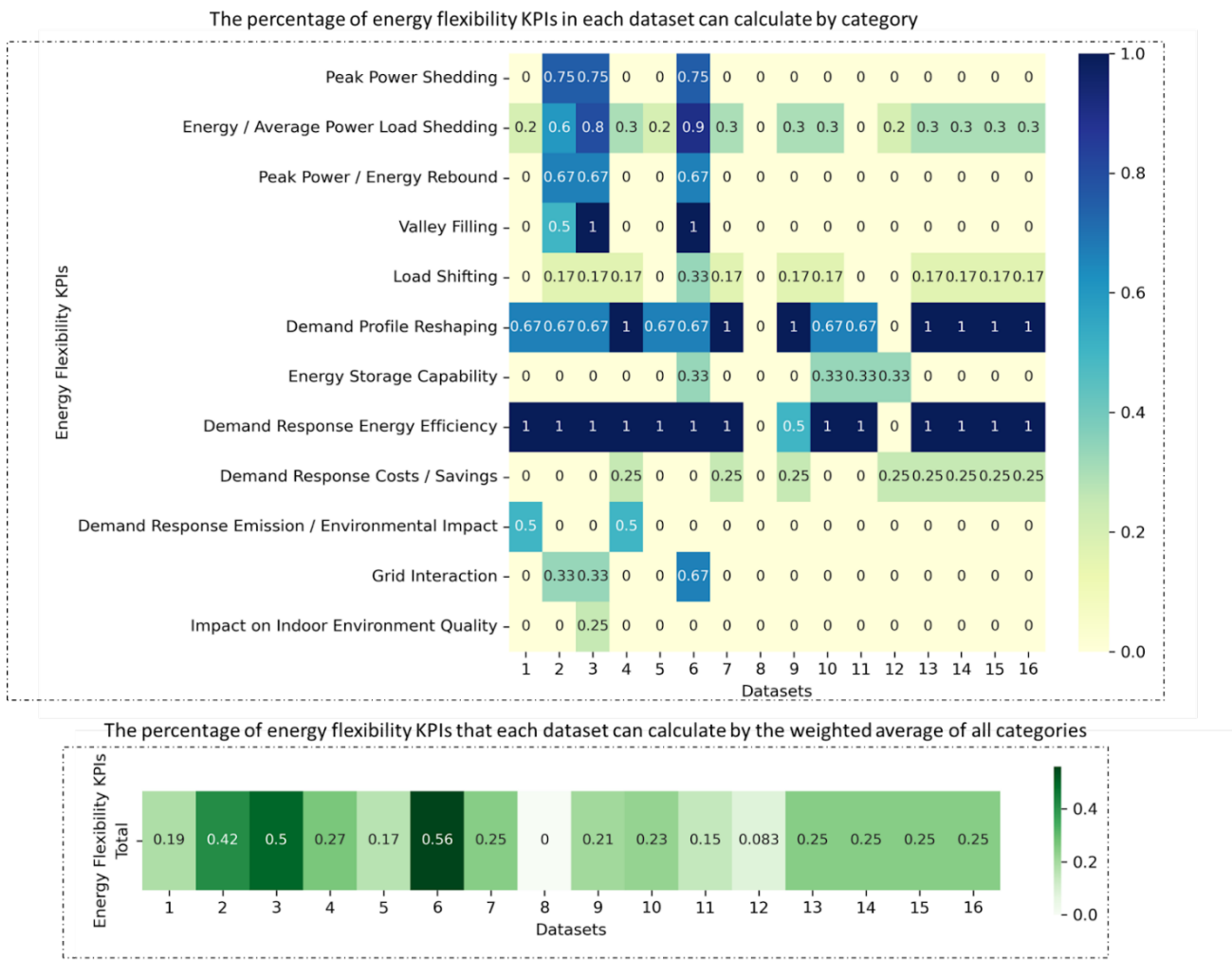

***Figure 12:*** *Heatmap contrasting the datasets' variables that are available with the ones needed to calculate each group of KPIs. The x-axis represents the 16 datasets that were collected, and the y-axis shows the 12 different categories of energy flexibility KPIs that were identified in this paper. The top and bottom subplots show the percentage of KPIs that*

*each dataset can calculate. The top one considers each category, while the bottom one considers a weighted average of all categories based on their respective number of KPIs.*

Six datasets are described in scientific publications or white papers. Most of the datasets are or are about to be open access, with two restricted access and two unspecified. However, to date, only four of the datasets are already published and available to download. The remaining datasets are either part of ongoing projects (the data publication is foreseen in 2023) or part of completed projects for which access to the data can be granted by contacting the research team.

*5.3. Use case analysis*

Despite the limited output size of the data collection process, it is still feasible to infer a number of trends regarding B2G application services for these case studies. As presented in *Figure 13*, load shifting and load shedding are the most common DSM control strategies. HVAC systems are the most commonly activated resource to deliver flexibility. The flexibility action is often triggered by temperature adjustments, uniquely applied in 19% of the datasets or within a mix of actions in the other 69%. The control strategy is often multi-objective; the most typical ones are to reduce peak demand, minimize energy costs, and maintain thermal comfort.

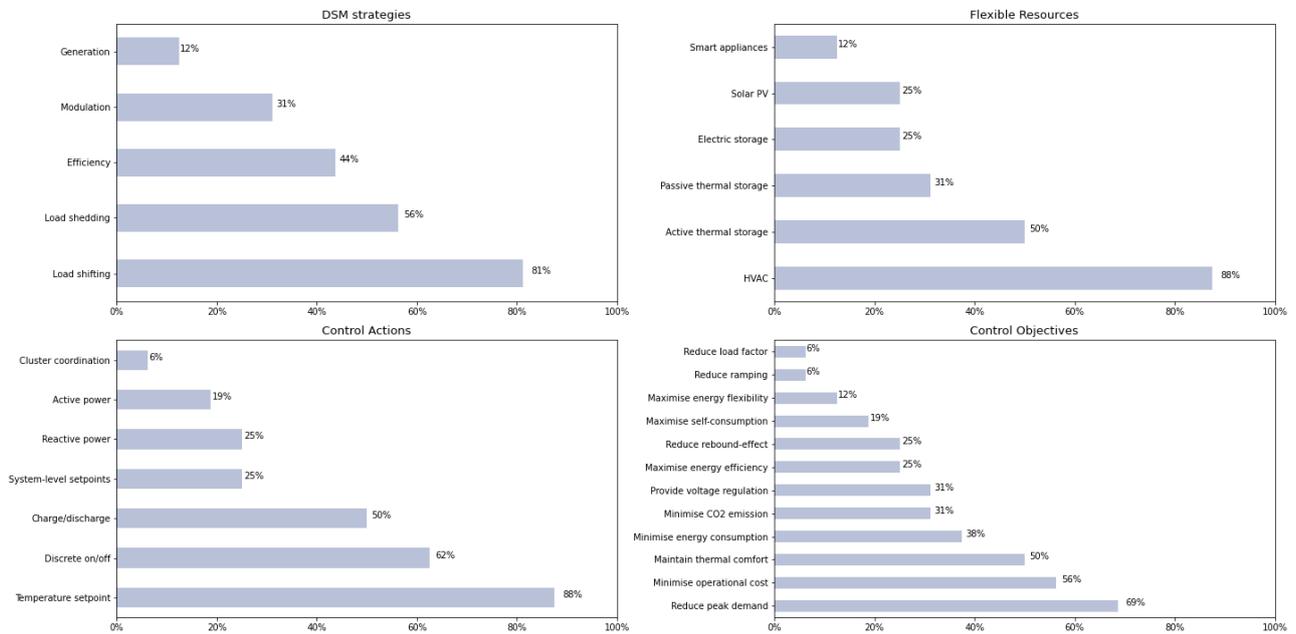

***Figure 13:*** *Overview of the collected datasets: DSM strategies and flexible resources (top), control actions and control objectives (bottom).*

More in-depth assessment can be made by segmenting the dataset information into instances based on five of their major features: DSM strategy, building sector, grid sector, flexible resources, and control actions (see *Figure 14*). The first feature confirms the clear preference for solutions related to load shifting and load shedding strategies. The grid sector criteria show that the vast majority of implemented solutions focus on the electricity grid, with only a small number dedicated to district heating. This is to be expected due to the low utilization of district heating networks worldwide [93]. Most strategy instances have been applied to commercial buildings, underlining HVAC systems as their most preferred target flexibility resource. This is strongly supported by the high share of HVAC systems in terms of energy demand, accounting for 38% in the building sector [94]. Lastly, for control action features, temperature control and on/off regulation are the most common, which is a direct consequence of the prevalence of HVAC-oriented solutions.

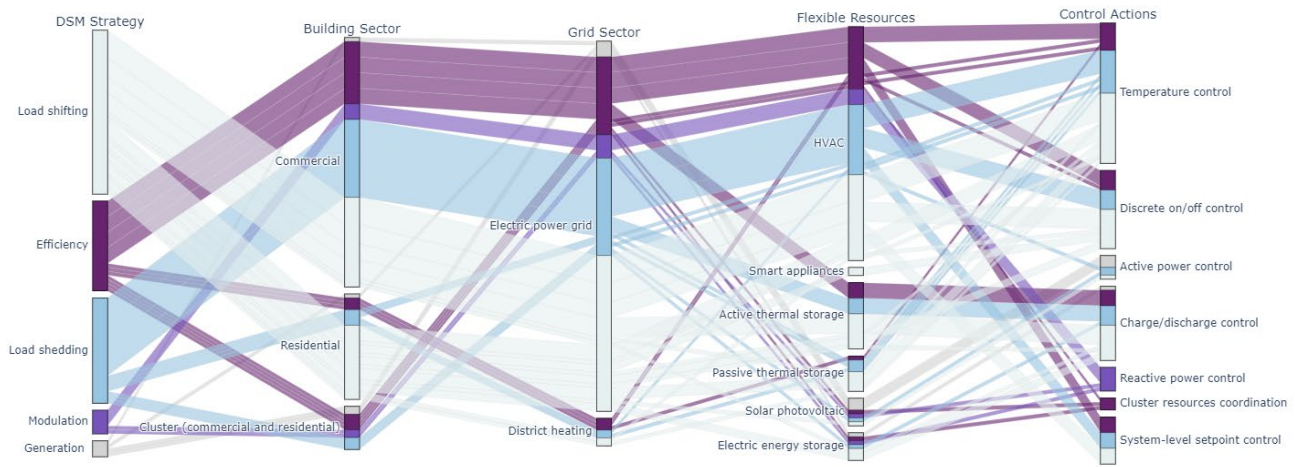

***Figure 14:*** *Interaction between DSM strategies, building sectors, grid sectors, flexible resources, and control actions within the analyzed datasets. The colour scale differentiates between the different DSM strategies. The thickness of the line connectors between each column represents the proportion of datasets from the previous category to the next one.*

A more comprehensive analysis was made by considering the interaction between the features presented in *Figure 14* (e.g., the relation between the flexible resources and proposed control actions). For HVAC systems, the control actions were based on on/off regulation and modulation-type controls such as output setpoint or internal system-level setpoint and active/reactive power control (e.g., fan modulation). In most cases, the HVAC-based strategies leverage the passive thermal inertia of the building (i.e., the thermal mass of the indoor environment) and active thermal storage systems via HVAC setpoints and charge/discharge adjustments. Active/reactive power and charge/discharge control actions are also used for solar PV and electrical energy storage, which are typically implemented together in these use cases. For a combination between HVAC, solar PV, and electrical storage, a coordinating action of cluster resources was applied for one of the study cases, aiming to increase the overall efficiency. Finally, for the single-use case taking into consideration smart appliances, discrete on/off regulation was implemented.

An important limitation of this analysis is that the DSM strategy classification was collected as a survey input for the majority of the datasets, and it is likely that different research teams have slightly different definitions for them, which could result in misclassifications.

*5.4. Limitations and future work related to dataset collection*

The limited number of datasets analyzed in this section may have hindered the potential for more robust conclusions of current B2G service applications. Although many relevant studies have produced simulation or monitoring data to test their hypothesis, getting access to these datasets is highly problematic or, in many instances, not possible at all. As indicated by the low reply rate of the contacted research teams (19%), the main obstacle is getting a response from the research teams performing these investigations. However, due to turnover in research institutions, the contact information indicated in studies can be outdated. Concerns over confidentiality created a second obstacle: despite the existence of datasets, many could not be released publicly, not even through confidentiality of restricted use agreements nor after data anonymization processes. The open-access data publication of case study projects must be significantly increased and planned early on during the design of these projects. The third significant barrier is the lack of available human resources within case study projects to format, curate, and correctly document datasets so other research teams can use them.

The analyzed datasets were originally collected with the intention of using them for testing and benchmarking the energy flexibility KPIs reviewed in this paper. Due to the heterogeneous formats and different use of these datasets, an investigation of data quality procedures (such as cleaning and filtering) will be needed [95][96][97][98][99]. The core focus of future work will be to establish a framework for testing KPIs based on the high and uniform quality of the datasets, which will serve as a first step towards standardizing the benchmarking in energy flexibility research.

## 6. Discussion

The review of 87 articles and technical reports reveals that 48 core energy flexibility KPIs and 29 more generic building performance KPIs have been used with data-driven approaches to asses B2G services. These KPIs focus on demand-side energy management performance in the operational phase of buildings and can be categorized using the hierarchical structure shown in *Figure 1*. Broadly defined, one group of KPIs can be calculated directly from a building's historical monitored sensor and meter data, while the other group of KPIs needs a data-driven method to estimate the baseline energy demand profile.

The three most commonly used data-driven energy flexibility KPIs, grouped based on whether a baseline is required or not, were identified and listed in *Table 3*. These KPIs focus on (1) the performance of load shifting, (2) energy cost reduction, and (3) peak demand reduction, which directly benefit building owners and grid operators.

This review makes contributions to the existing body of knowledge on building energy flexibility KPIs in: (1) a comprehensive analysis and categorization of existing data-driven energy flexibility KPIs, (2) an in-depth understanding of use cases and stakeholders of the KPIs, (3) guidelines to choose adequate KPIs, and (4) a wide survey and inventory of building performance datasets that can be used to quantify energy flexibility KPIs using data-driven approaches.

Major research gaps in data-driven energy flexibility KPIs were also identified, and these point to future research opportunities, including the following:

1) **Data-driven methodology development:** It was found in the review that more than 80% of the energy flexibility KPIs are baseline-required. However, acquiring both baseline and flexible mode data is still a challenge for real buildings in operation. Most existing data-driven methodologies were not specifically designed for demand response. Future efforts are needed to develop data-driven methodologies that can evaluate different energy flexibility strategies (e.g., energy storage, multi-energy systems) under different scenarios (e.g., weather conditions, occupancy).
2) **Baseline-free energy flexibility KPI development:** Although baseline-free KPIs are relatively easy to calculate, they only make up less than 20% of the energy flexibility KPIs. Therefore, future opportunities exist for developing such KPIs that capture different energy flexibility scenarios and performance goals.
3) **Energy flexibility KPIs to support real applications:** It was found from reviewing the current scientific literature on the topic that most studies focused on technical perspectives of building energy flexibility in experimental settings. However, there are still many barriers in real applications, like standardized procedures to measure and verify building energy flexibility. Future energy flexibility KPI development should consider factors beyond the engineering perspectives, such as flexibility markets, building occupants' behavior, acceptability and feedback, utility programs, building codes and standards, and integration of building energy flexibility KPIs into other operational metrics such as energy use, energy cost, and carbon emissions for holistic building performance assessment.
4) **Dataset collection for energy flexibility assessment:** Open datasets with proper descriptions for energy flexibility assessments are still very limited. Most datasets were not designed or collected with the energy flexibility quantification as the objective from the beginning, leading to missing data points or unmatched spatial-temporal resolutions. The presented collection of datasets of buildings performing B2G services in this study is preliminary and will be continued within the IEA EBC Annex 81 activities. These datasets will be reviewed and analyzed in depth and used to test and investigate further the different energy flexibility KPIs. These collected datasets also can serve as showcase examples of what buildings can achieve in terms of B2G services in various conditions.

## 7. Conclusions

This paper presents a holistic review of data-driven energy flexibility KPIs for operational buildings. An initial set of 156 articles and technical reports were identified, of which 87 were selected for in-depth review. A suite of 48 core energy flexibility KPIs and 29 generic building performance KPIs were reviewed in terms of their use cases, stakeholders, performance goals, applicable temporal and spatial scales, data requirements, and calculation complexity. Depending on the data-driven methodology, the KPIs were divided into two groups: baseline-required KPIs (39), which require building performance data in both flexible and reference scenarios, and baseline-free KPIs (9), which could be calculated without a reference scenario. A brief summary of data-driven baseline generation methodologies is included for baseline-required KPIs. A multi-step process is proposed to facilitate KPI selections.

Data is the fundamental ingredient of data-driven energy flexibility KPIs. To examine which and how existing datasets can support data-driven energy flexibility assessments, a survey of datasets related to B2G applications was conducted. Initially, 330 datasets or dataset descriptions were identified as potentially being linked to research studies and pilot projects that could include demand response, demand-side management, or energy flexibility strategies. However, only 16 (4.8%) of the datasets were found to be within the scope and have a proper description and data availability. That subset was then analyzed with regard to the energy flexibility KPIs and use cases that each dataset can support. The review makes contributions to the existing body of knowledge on building energy flexibility KPIs and provides several insights into research areas that require further attention.

Finally, this current work recognises the importance of user characteristics in the context of data-driven energy flexibility KPIs. In this article, it has been acknowledged that while data-driven solutions may not rely on preexisting detailed models or prior knowledge of building characteristics, including occupants' behavior and interaction is of critical importance. Therefore, incorporating occupant comfort and acceptance of reduced comfort into KPI development can provide valuable insights and improve the performance of B2G services, and it will be the object of future work.

**Acknowledgements**

This work has been carried out within the framework of the International Energy Agency (IEA) Energy in Buildings and Communities (EBC) Annex 81: "Data-Driven Smart Buildings" (https://annex81.iea-ebc.org). The authors would like to gratefully acknowledge the IEA EBC Annex 81 for providing an excellent research network and thus enabling fruitful collaborative research studies like the present one.

Lawrence Berkeley National Laboratory's (LBNL) effort was supported by the Assistant Secretary for Energy Efficiency and Renewable Energy, Building Technologies Office, of the U.S. Department of Energy under Contract No. DE-AC02-05CH11231.

University College Dublin's (UCD) and CARTIF's effort was supported by: (a) the CBIM-ETN funded by the European Union's Horizon 2020 research and innovation programme under the Marie Skłodowska-Curie grant agreement No 860555 and, (b) the NeXSyS project under the auspices of Science Foundation Ireland (SFI) Grant 21/SPP/3756.

Texas A&M University's (TAMU) effort was supported by: 1) NSF project # 2050509 "PFI-RP: Data-Driven Services for High Performance and Sustainable Buildings." and 2) the Building Technologies Office at the U.S. Department of Energy through the Emerging Technologies program under award number DE-EE0009150.

Syracuse University's (SU) effort was supported by the U.S. National Science Foundation (Award No. 1949372).

Aalborg University's (AAU, Denmark) effort was supported by the IEA-EBC Annex 81 Participation project funded by The Energy Technology Development and Demonstration Programme - EUDP (Case no. 64019-0539).

DEWA R&D Center's effort was supported and funded by the Dubai Electricity and Water Authority.

Concordia University's effort was supported by funding from the Canada Excellence Research Chairs Program and the Tri-Agency Institutional Program Secretariat (Grant CERC-2018-00005), the Natural Sciences and Engineering Research Council of Canada (Discover Grant RGPIN 2020-06804), and the Fonds de recherche du Québec: Nature et technologies (FRQNT) Doctoral Research Scholarship.

**Conflict of interest**

The authors declare that they do not have competing financial interests, personal relationships or any kind of known conflict of interest that could have influenced the work reported in this article.

## Appendices

The different appendices and supplementary material created for this study can be found on the GitHub dedicated to the IEA EBC Annex 81 (Data-Driven Smart Buildings) – SubTask C3 (Building to Grid Applications): https://annex-81-c3.github.io/data-driven-KPIs-review/.

*Appendix A*

*Table A1* is a complete list of the 87 selected reviewed articles with a brief summary of the use cases, applicable building sectors, flexibility resources, quantification methods, and potential stakeholders for each paper [22][23][24][25][26][27][28][29][30][31][32][33][34][35][38][39][40][41][42][45][60][61][62][63][64][65][66][67][69][70][71][72][73][74][76][77][78][80][81][100][101][102][103][104][105][106][107][108][109][110][111][112][113][114][115][116][117][118][119][120][121][122][123][124][125][126][127][128][129][130][131][132][133][134][135][136][137][138][139][140][141][142][143][144][145][146][147]. *Table A1* can be found in the following Google spreadsheet document: https://docs.google.com/spreadsheets/d/1BYvYF_kVScc9upolPzEZHnIsrvBP2N2_1fS-nBxO2nI/edit#gid=1335917425.

*Table A2* is a complete list of the 48 collected data-driven energy flexibility KPIs and the 29 generic building KPIs associated with energy flexibility studies in the reviewed publications. *Table A2* can be found in the following Google spreadsheet document: https://docs.google.com/spreadsheets/d/1BYvYF_kVScc9upolPzEZHnIsrvBP2N2_1fS-nBxO2nI/edit#gid=978094966.

The code/script used to analyze the data of *Table A1* and *Table A2* can be found in the following Google Colaboratory notebook: https://colab.research.google.com/drive/1gbz13aGcwLCQLryAQufPZywPOp-QQmW0.

*Appendix B*

*Table B1* is a complete list of the 16 identified available datasets of buildings performing DR with detailed information about their respective study and a link to download the datasets when the latter are available. *Table B1* can be found in the following Google spreadsheet document: https://docs.google.com/spreadsheets/d/1oeuAFI0595vohN7Fvo8apNX_4faoa-SyQUJTlcVQ0Bo/edit?usp=sharing.

The code/script used to analyze the data of *Table B1* can be found in the following Google Colaboratory notebook: https://colab.research.google.com/drive/1xg7Z6nkgdF7o4sU7VZSM4zHMIhxiCZEb?usp=sharing.